\documentclass[reprint,floatfix,nofootinbib,preprintnumbers,amsmath,amssymb,superscriptaddress,pra]{revtex4-2}

\usepackage{graphicx}
\usepackage[margin=1in]{geometry}
\usepackage{amssymb}

\usepackage{booktabs}
\usepackage{tabulary}
\usepackage{rotating}
\usepackage{enumitem}
\usepackage[breaklinks=true]{hyperref}
\usepackage{xcolor}
\usepackage{siunitx}

\newcommand{\code}[1]{\texttt{#1}}

\begin{document}

\title{A high-performance and portable implementation of the SISSO method for CPUs and GPUs}

\author{Sebastian Eibl}
\affiliation{Max Planck Computing and Data Facility, 285748 Garching, Germany}
\author{Yi Yao}
\affiliation{The NOMAD Laboratory at the FHI of the Max Planck Society, D-14195 Berlin, Germany}
\author{Matthias Scheffler}
\affiliation{The NOMAD Laboratory at the FHI of the Max Planck Society, D-14195 Berlin, Germany}
\author{Markus Rampp}
\affiliation{Max Planck Computing and Data Facility, 285748 Garching, Germany}
\author{Luca M. Ghiringhelli}
\affiliation{The NOMAD Laboratory at the FHI of the Max Planck Society, D-14195 Berlin, Germany}
\affiliation{Department of Materials Science and Engineering, Friedrich-Alexander-Universität Erlangen-Nurenberg, Germany}
\author{Thomas A. R. Purcell}
\affiliation{The NOMAD Laboratory at the FHI of the Max Planck Society, D-14195 Berlin, Germany}
\affiliation{Department of Chemistry and Biochemistry University of Arizona, Tucson, Arizona, USA}

\begin{abstract}
SISSO (sure-independence screening and sparsifying operator) is an artificial intelligence (AI) method based on symbolic regression and compressed sensing widely used in materials science research. SISSO++ is its C++ implementation that employs MPI and OpenMP for parallelization, rendering it well-suited for high-performance computing (HPC) environments.
As heterogeneous hardware becomes mainstream in the HPC and AI fields, we chose to port the SISSO++ code to GPUs using the Kokkos performance-portable library. Kokkos allows us to maintain a single codebase for both Nvidia and AMD GPUs, significantly reducing the maintenance effort.
In this work, we summarize the necessary code changes we did to achieve hardware and performance portability. This is accompanied by performance benchmarks on Nvidia and AMD GPUs. We demonstrate the speedups obtained from using GPUs across the three most time-consuming parts of our code.
\end{abstract}

\date{\today}
\maketitle

\section{Introduction}
A look at the TOP500 list of the most powerful supercomputers~\cite{TOP500} shows that hardware has diversified over the last years. CPUs from Intel and AMD as well as ARM based ones together with GPUs from AMD, Nvidia and Intel are available for the users. It is also very likely to see this trend continue in the future with the rise of integrated chips like Grace Hopper (Nvidia) or MI300A (AMD) and accelerators tailored for specific tasks.

This complicates the work of software developers who need to enable their codes to run on a broad range of current, and ideally also future, architectures, particularly since the available vendor-native programming models are usually restricted to only a single hardware platform of the corresponding vendor.
Some applications maintain multiple versions of the code implemented with the different native programming models in order to fully utilize the specific hardware by taking advantage of more detailed controls provided by the native models (e.g. Nvidia's CUDA). However, this approach demands significant effort from developers in both coding and maintaining these separate versions.

To address this challenge, architecture-independent programming models have emerged as alternatives \cite[cf.][]{davis2024}. They can be broadly categorized into directive-based and library-based approaches. Examples for directive-based approaches are OpenMP and OpenACC. These approaches facilitate reusing existing CPU-based code and incrementally porting performance-critical sections by offloading them to accelerators. Several success stories have demonstrated the effectiveness of these models \cite{marowka2022}, but, depending of the programming language and the hardware platform, compilers are on different --- and partly still insufficient --- levels of maturity~\cite{herten2023}.

Library-based approaches like std::par~\cite{Latt2021}, RAJA~\cite{Beckingsale2019} or Kokkos~\cite{Trott2022} have also helped developers achieve vendor and platform independence. They provide high-level parallel abstractions for common parallel patterns like `for\_each', `reduce', and `scan' together with memory management utilities. These frameworks map such abstractions to hardware-specific backends utilizing platform-native programming models. This hides the maintenance burden from the application developer by offloading it to the shoulders of a (potentially) larger and more specialized community of developers backing these projects.

After having decided on the programming model, one often needs to adapt both the algorithms and the data structures in order to get the best performance on different hardware architectures. Changes to algorithms originally implemented for CPUs become necessary because GPUs are optimized for executing thousands of operations in parallel using very many, but comparably weak compute units, whereas CPUs are designed for efficient single-threaded performance and complex task handling with fewer, but more powerful cores. Additionally, optimal memory access patterns differ between CPUs and GPUs which often requires to also adapt the data structures. As a rule of thumb, CPUs perform better with an array-of-structs (AoS) data layout, while GPUs prefer struct-of-arrays (SoA) \cite[e.g.][]{ruzicka2024}. This must be taken into account when designing the most important and frequently used data structures.

In this work we describe our experiences with integrating the Kokkos library into our C++ code SISSO++ \cite{Purcell2022,Purcell2023} in order to extend the existing CPU-based code base for the use of GPUs. SISSO++ is licensed under the Apache 2 Licence and can be obtained free of charge from GitLab\footnote{\url{https://gitlab.com/sissopp_developers/sissopp}}. SISSO++ implements the SISSO method (see the next section), an artificial intelligence method based on symbolic regression and compressed sensing, which is widely used in materials science research. The method is explained in detail in the next section. It is computationally expensive and applications in materials research require significant high-performance computing resources, making the efficient and scalable use of modern, GPU-accelerated platforms a must. To this end, Kokkos helps manage parallelism and memory access patterns, allowing us to make the SISSO++ code both architecture-independent and performance-portable. In the following sections, we will first introduce the SISSO method, then describe how we used Kokkos to port the three most time-consuming steps of our algorithm. We will also present two test cases used to benchmark our porting efforts, along with the performance results obtained on CPU-only and GPU-accelerated platforms of different vendors and hardware generations, namely Intel Xeon IceLake-SP CPUs, together with the GPU models, Nvidia A100, AMD MI250, and the new AMD MI300A.

\section{SISSO}

\begin{figure*}
   \centering
   \includegraphics{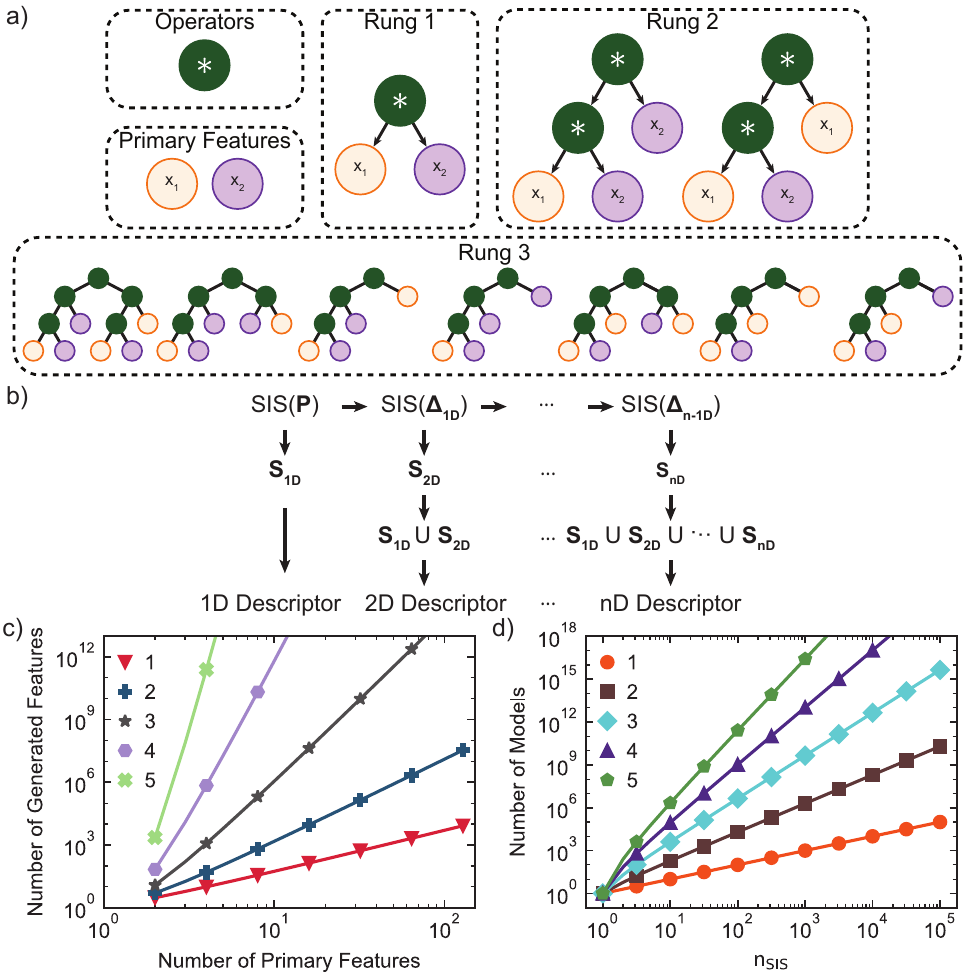}
   \caption{
   Illustration of the SISSO method. a) An example of the feature creation step of SISSO with two primary features $x_1$ (light orange) and $x_2$ (purple) and the multiplication operator (dark green). b) A flowchart of the algorithm used for the descriptor identification step. c) A comparison of the maximum number of features generated for the multiplication operator with respect to the number of primary features for rungs 1 (red triangle), 2 (blue plus), 3 (gray star), 4 (lavender hexagon), and 5 (green x). d) A comparison of the number of models evaluated against the size of the SIS subspace for a 1 (orange circle), 2 (brown square), 3 (blue diamond), 4 (purple triangles), and 5 (green pentagons) dimensional model. }
   \label{fig:method_cartoon}
\end{figure*}

The sure-independence screening and sparsifying operator (SISSO) approach is an artificial intelligence (AI) method that combines compressed sensing and symbolic regression~\cite{Ouyang2017,Ouyang2019,Purcell2022,Purcell2023}. SISSO offers significant advantages by efficiently reducing high-dimensional data to a small set of relevant features, creating interpretable and compact models through symbolic regression. It excels in producing accurate, sparse models that are computationally efficient and easy to understand, making it ideal for materials science research. Additionally, it can provide low-dimensional maps for visualizing materials space.

The goal of SISSO is to find the best low-dimensional analytical descriptor for a given target property, $P$.
SISSO is broken up into two stages, feature creation and descriptor identification, each illustrated in Figure~\ref{fig:method_cartoon}a and \ref{fig:method_cartoon}b, respectively.
In the feature creation step, the algorithm starts with a set of user-defined analytical operators (e.g. addition, subtraction, multiplication, etc.) and primary features. The primary features are vectors containing a specific property of all available samples (e.g. radius, charge, etc.). From there, a larger pool of candidate descriptors (i.e. a set of features that can approximate the target property by linear combination) in rungs are created by applying the operators onto primary feature and previously generated equations, while preserving unit consistency and ensuring the domain of the operations is respected~\cite{Purcell2023}.
As the rung of each candidate descriptor (height of the binary expression trees in Figure~\ref{fig:method_cartoon}a) increases, the number of possible equations increases combinatorially for all binary operators, as illustrated for the multiplication operator in Figure~\ref{fig:method_cartoon}c.
This typically results in billions or trillions of functions that build the offered feature space.
The descriptor identification step is broken up into two main components, sure-independence screening (SIS)~\cite{Fan2008} to create a subspace of selected descriptors, $\mathcal{S}$ and a $\ell_0$-regularized optimization over $\mathcal{S}$, as shown in Figure~\ref{fig:method_cartoon}b~\cite{Ouyang2017,Ouyang2019}.
For each dimension new descriptors are added to $\mathcal{S}$ based on a projection score against either $\mathbf{P}$ for the first dimension, or the residual of the previously generated models, $\mathbf{\Delta_{n-1D}}$.
The need for the SIS step is highlighted in Figure~\ref{fig:method_cartoon}d as the workload of $\ell_0$-regularization scales combinatorially with the number of candidate descriptors used in the final linear model. While the one-dimensional model found by SISSO is the exact $\ell_0$-regularized one , finding a two-dimensional model can be impractical if searching over the complete feature space.

\subsection{Starting point}
The starting point of the code development reported in this paper is the last published version of SISSO++~\cite{Purcell2022}. This version implements a hybrid MPI+OpenMP parallelization. All compute intensive parts use the Intel MKL library to achieve good performance on x86 CPUs. However, this version can only use CPUs.

\subsection{Kokkos}
We decided to use the Kokkos framework for porting the SISSO++ code to modern GPU-based platforms in a hardware-agnostic way.
Our choice of Kokkos is motivated by first-hand experience gained in our own group and close collaborators \cite{artigues2020,schikd2024,ruzicka2024}, together with observation of the literature \cite[e.g.][]{evans2022,marowka2022,lesur2023,roussel2024}, leading to the perception that
Kokkos is a mature and future-proof portability framework for HPC applications, that it is actively developed and supported, with new versions being released regularly. The current version, as of writing this paper, is \code{4.4.01}.

Kokkos provides the necessary functionality for writing hardware-agnostic code. Specifically, for parallelizing code, we use three parallel constructs, namely \code{Kokkos::parallel\_for} (for-each), \code{Kokkos::parallel\_reduce} (reduce) and \code{Kokkos::parallel\_scan} (scan).
In addition, we take advantage of portable data structures provided by Kokkos: \code{Kokkos::View} is a multidimensional array abstraction for memory located on either host, device or in unified shared memory. Moreover, \code{Kokkos::View} can be allocated in shared memory on GPUs.
These features are sufficient to express all parts of our code that are supposed to be executed on the GPU.

\subsection{Feature generation and SIS}
\begin{figure*}
    \centering
    \includegraphics[scale=0.5]{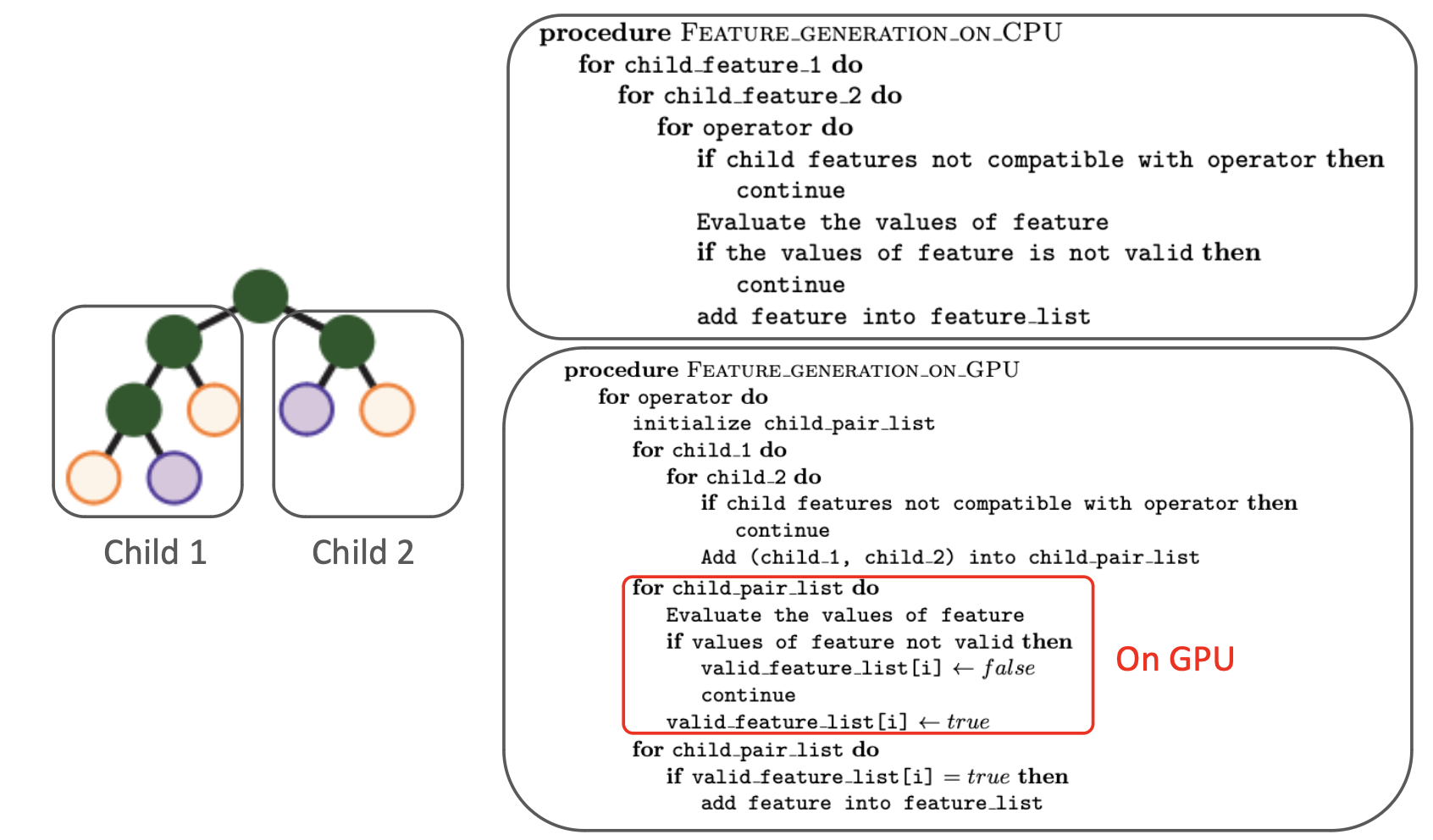}
    \caption{CPU and GPU algorithms for feature generation}
    \label{fig:feature_generation_algorithm}
\end{figure*}

Feature generation and SIS are the steps needed to generate candidate expressions, for the one-dimensional $\ell_0$-regularized model and the subsequent addition of more dimensions.
During the feature generation phase, as depicted in Figure \ref{fig:feature_generation_algorithm} a) and b), higher-rung features are constructed by combining lower-rung features using arithmetic operators. We refer to these lower-rung features as ``child features".

However, not all combinations of child features are valid. We have established rules to eliminate duplicate and ill-defined features, based on either the properties of the child features or the values of the generated feature. For instance, we avoid constructing features that contain zeros in its second child for the divisor operator. Additionally, we enforce rules to ensure that all generated feature values adhere to predefined lower and upper bounds, are not the identical, and do not contain ``not a number" (NaN) values. Detailed explanations of these rules rules are given in prior work\cite{Purcell2023}.

The rules based on child features can prevent unnecessary calculations of new feature values, while the other rules require the evaluation of the new feature values to ensure validity.

A comparison between the feature generation algorithms on CPU and GPU is shown in Figure. \ref{fig:feature_generation_algorithm}.
The CPU implementation involves straightforward nesting of three loops over \textit{child\_feature\_1}, \textit{child\_feature\_2}, and types of \textit{operator}.
In the GPU implementation, we reconstructed the loop order so that the types of operator form the outer loop.
This allows us to process features with the same operators concurrently.
For the rules based on child features, we still evaluate them on the CPU to generate a list of potential child feature pairs for each operator.
These potential pairs along with the values of child features are transferred to GPU.
Using the Kokkos team policy, we perform the value evaluations and apply the value-based rule checks on the GPU, leveraging its shared cache memory.
The GPU execution produces a validity list for the potential pairs, based on which we construct valid features on the CPU. Due to high memory requirement, we opt not to store values of features for the last rung in memory.
Saving features themselves can be beneficial as they can be reused across all the dimensions. However, the large number of features can lead to prohibitive memory requirements. As an alternative, we offer another memory management strategy that generates features on-the-fly for SIS in each dimension, thus containing memory consumption.

For SIS, we evaluate the Pearson correlation coefficients ($r$) of all the generated features ($x$) to the target property or the residual of the previous dimensions ($y$) according to the formula
\begin{equation}
    r=\frac{\sum\left(x_i-\bar{x}\right)\left(y_i-\bar{y}\right)}{\sqrt{\sum\left(x_i-\bar{x}\right)^2 \sum\left(y_i-\bar{y}\right)^2}} \;.
\end{equation}

Since we chose not to store the values of the last rung features, we re-evaluate them before calculating the Pearson correlation coefficients with the target property or residual. This re-evaluation and the subsequent Pearson correlation calculation are performed consecutively on the GPU, where we found that a batch size of 50 to 100 million provides an optimal balance between speed and memory usage. Here batch size refers to the number of features passed to the GPU over one iteration. The resulting Pearson correlation coefficients are then transferred back to CPU, where they are used to rank the features and select the top candidates for the subsequent $\ell_0$ regularization.

\subsection{$\ell_0$-regularization}
The input for this algorithm is a set of $m$ top-rated features produced by the previous SIS step. Its purpose is to find the best descriptor formed by tuples of $n$ features. This is done by scoring every possible descriptor and returning the best one. For the scoring, a least-squares approximation of all features of the descriptor to the target property is calculated, and the mean squared deviation is used as a score.
The individual steps executed for each set of features are as follows:
    \begin{itemize}[nosep]
        \item assemble the descriptor matrix in scratch space
        \item QR factorization of the matrix using Housholder reflections
        \item calculate least squares fit
        \item calculate scores as mean squared deviation
    \end{itemize}

\noindent
Our porting and further improvements of this algorithm are based on the concepts of loop fusion and batched execution, and take advantage of the concurrent usage of both, CPU and GPU resources as well as automatic tuning of offloading parameters and the option to run the program with lower arithmetic precision. Each of these ingredients is detailed in the following.

Since individual linear systems are small, the matrix and the right-hand side fit into the GPU's shared memory for the whole computation. We avoid unnecessary copies to and from shared memory by fusing all steps into one Kokkos kernel. Fusing all steps somewhat deteriorates readability of the code, but the performance gain compared to several calls to individual kernels is significant.

The typical size of a feature vector is around $1\,000$ and the dimension $n$ is smaller than 10. This leads to small and skinny matrices for the least-squares solver. For CPUs, this is not an issue. However, initial tests showed that today's powerful GPUs cannot be fully utilized by scoring each descriptor individually. We, therefore, employ a batched solver on GPUs that handles multiple descriptors at the same time which means that one GPU thread scores one descriptor. The exact size of a batch can be configured via the configuration script. For modern GPUs, batch sizes should exceed $65\,536$ in order to fully saturate the available resources.

Since scoring different descriptors has no interdependencies, parallelization over the descriptors is trivial, thus allowing efficient execution on GPUs, and enabling the concurrent use of both, the GPUs and the host CPUs. We use OpenMP tasks for scheduling batches onto CPU threads and asynchronously offloading others to the GPU. For each batch, one OpenMP task is created. The task scheduling system of OpenMP assigns waiting tasks to idling CPU threads. This already allows us to utilize the available CPU resources. The GPU offloading is achieved by using one CPU thread to offload its batches to GPU. Depending on the hardware in use, it can be beneficial to oversubscribe the number of CPU threads by one to not waste a CPU core only for offloading.

Depending on the size of the data-set and the hardware in use, a different launch configuration (in Kokkos terminology the team size) yields the best performance. By applying auto-tuning we ensure to use a good configuration for the setup at hand. The auto-tuning uses the first batch during each run to evaluate the time-to-solution of a predefined set of launch configurations. The best one is chosen and used for the rest of the batches. In particular the number of threads per block plays an important role as the native warp size is different between Nvidia and AMD hardware (32 and 64, respectively). The auto-tuning itself introduces some overhead for computing the first few batches, but in practice this amounts to only a few seconds which is negligible compared to the total runtime of a large number of batches. Therefore, auto-tuning is always applied as the overall savings can be significant.

While restructuring the code base we also introduced the option to select the floating point precision to be either double precision (FP64), which was the default so far, or the newly introduced single precision (FP32) for which modern datacenter GPUs typically offer at least twice the theoretical peak floating-point performance compared to double precision. Some parts of SISSO++ can now profit from such performance boost, provided the lower precision is tolerable for the given application. Overall the performance gain is less than a factor of two due to non-numerical operations.

\section{Performance and Portability}

\begin{table*}
    \begin{tabular}{lccc}
    \toprule
         & A100 & MI250 & MI300A \\
    \midrule
         CPU & Intel Xeon IceLake Platinum 8360Y & AMD EPYC 7643 & MI300A\\
         cores & 2x36 & 2x48 & 2x24\\
    \midrule
         GPU & 4x Nvidia A100 SXM4-40GB & 4x AMD Instinct MI250 & 2x AMD Instinct MI300A\\
         peak performance FP64 & 4 x 9.7 TFLOP/s & 4 x 45.3 TFLOP/s & 2 x 61.3 TFLOP/s \\
         peak performance FP32 & 4 x 19.5 TFLOP/s & 4 x 45.3 TFLOP/s & 2 x 122.6 TFLOP/s \\
         peak memory bandwidth & 4 x 1.6 TB/s & 4 x 3.2 TB/s & 2 x 5.3 TB/s \\
    \midrule
         host compiler & icpc 2021.7.1 & hipcc & hipcc \\
         SDK & CUDA 12.2 & ROCm 6.2 & ROCm 6.2 \\
         device compiler & nvcc & hipcc & hipcc \\
         Kokkos version & 4.4.01 & 4.4.01 & 4.4.01 \\
    \bottomrule
    \end{tabular}
    \caption{Basic performance characteristics of the heterogeneous Nvidia and AMD systems used for our benchmarks. We use both GCDs of the MI250 GPU.}
    \label{tab:hardware}
\end{table*}

In this section we assess the performance and portability of our new code on Nvidia and AMD hardware. The hardware specifications of the machines are summarized in \autoref{tab:hardware}. For benchmarking on Nvidia hardware we use the HPC cluster \textit{raven} at the Max Planck Computing and Data Facility MPCDF). For development and initial profiling of the code for AMD GPUs we received access to a pre-release MI300A in the AMD Accelerator Cloud (AAC). The final performance numbers on MI300A presented here were obtained  on the \textit{viper} cluster also located at the MPCDF.
We assess the performance in different configurations:
\begin{itemize}[nosep]
    \item \textbf{CPU}: CPU-only reference configuration using all CPU cores available. This configuration uses the highly optimized Intel MKL library for all algebraic operations.
    \item \textbf{GPU-FP64}: GPU-accelerated configurations, where all available GPUs per node are used (GPU). The expensive parts are computed in double precision.
    \item \textbf{GPU-FP32}: GPU-accelerated configurations, where all available GPUs per node are used (GPU). The expensive parts are computed in single precision.
    \item \textbf{CPU+GPU-FP64}: All GPU resources and all CPU cores are used concurrently. The expensive parts are computed in double precision.
\end{itemize}

\noindent
The benchmarks reported here use one entire node of the respective cluster (or multiple nodes for the MPI scaling benchmarks).

\subsection{Test Cases}
In order to assess the performance of the new code we create two test cases, representative of a standard-use case and a large-data example, respectively.
For the standard use case, we use the thermal conductivity data-set published by Purcell, \textit{et al.}~\cite{Purcell2023a}, augmented by data from Knoop, \textit{et al.}~\cite{Knoop2023} in a multi-task SISSO configuration~\cite{Ouyang2019}.
Multi-task SISSO splits the data-set into multiple different sub-problems, tasks, and then tries to find an optimal model with the same descriptor matrix, but different coefficient matrices for each task.
For the large-data example we use a $2\,400$ material subset of the $3\,000$ compounds used in the NOMAD 2018 Kaggle competition describing the band gap energy of (Al$_x$In$_y$Ga$_{1-x-y}$)$_2$O$_3$ materials~\cite{Sutton2019}.
The definitions of the corresponding hyper-parameters are described in greater details in the subsequent Subsections~\ref{subsubsec:thermal_cond_test} and \ref{subsbusec:kaggle_test}. \autoref{tab:testcase_parameters} summarizes the most important parameters of the two test cases.

\begin{table*}
    \begin{tabular}{lrrr}
    \toprule
         & Thermal Conductivity & Kaggle Competition \\
    \midrule
        \# primary features & 17 & 12 \\
        data-set size & \num{156} & \num{2400} \\
        operators used & $+, -, *, \div, |x-y|$ & $+, -, *, \div, |x-y|$\\
                       & $\sqrt{x}, \sqrt[3]{x}, x^2, x^3, x^{-1}$ & $\sqrt{x}, \sqrt[3]{x}, x^2, x^3, x^{-1}$ \\
                       & $\log{x}, \exp{x}, \exp{-x}, |x|$ & $\exp{x}, $\\
        SIS subspace size & \num{2000} & \num{50000} \\
        \# evaluated descriptors in the final stage & \num{3587402000} & \num{4999950000} \\
    \bottomrule
    \end{tabular}
    \caption{Summary of the most important parameters for the two test cases.}
    \label{tab:testcase_parameters}
\end{table*}

\subsubsection{Thermal Conductivity Test Case}
\label{subsubsec:thermal_cond_test}
The thermal conductivity data-set consists of 75 experimental~\cite{Purcell2023a} and 57~\cite{Knoop2023} calculated thermal conductivity measurements. We separate these into two classes, experimental and calculated, to test the GPU-implementation for multi-task SISSO~\cite{Ouyang2019}.
This data set is further augmented by duplicating twelve rock-salt materials for both the experimental and aiGK tasks, leading to a total data set size of 156 materials.
We then use the same seventeen primary features used by \citet{Purcell2023a} as the primary features for this example.
All calculations use a maximum rung (the height of binary expression tree) of three, a maximum of three-dimensional descriptors, a SIS selection subspace size of $5\,000$, ten residuals per SIS iteration, a maximum absolute value for each candidate descriptor between 10$^{-5}$ and 10$^8$, and the addition, subtraction, multiplication, division, absolute difference, logarithm, exponential, negative exponential, absolute value, square root, cube root, square, cube, and inverse operators, leaving out the sin, cos and sixth power operators.
For this combination, $20\,820\,835\,000$ models are subjected to the final step of the $\ell_0$ regularization step.
This test case generates the last rung of features on-the-fly for each SIS iteration, and therefore can only use MPI+Kokkos (without OpenMP), because the CPU-only checks for the on-the-fly feature generation are not yet parallelized with OpenMP.
For on-the-fly calculations, the final rung of features are generated while calculating the SIS projection scores, instead of at the start of the runs.
This is done in cases where the final rung would be too large to fit into memory.

\subsubsection{Kaggle Competition Test Case}
\label{subsbusec:kaggle_test}
The Kaggle competition data set consists of $2\,400$ (Al$_x$In$_y$Ga$_{1-x-y}$)$_2$O$_3$ calculated band gap for different values of $x$ and $y$.
For this dataset we use twelve primary features: the six lattice parameters for each material; the values of $x$, $y$, and $1-x-y$; and the effective coordination number (ECN) of Al, Ga, and In.
ECN represents the effective number of nearest neighbor atoms surrounding a particular atom.
The SISSO-specific hyper-parameters were chosen to test the performance of the performance of SISSO on larger data sets, a descriptor dimension of two, a SIS subspace size of $50\,000$, ten residuals per SIS iteration, a maximum absolute value for each candidate descriptor between 10$^{-3}$ and 10$^5$, and all non-parameterized operations except sin, cos, log, negative exponential, absolute value, and sixth power.
With this combination, we expect to generate a total of \num{465242552} candidate descriptors and test $1\,249\,975\,000$ possible models during the last $\ell_0$-regularization step.
For GPU calculations the Kokkos team size was auto-tuned. The $\ell_0$ batch size used is $131\,072$, and the feature generation batch size is $\times10^{8}$.
For this example, all candidate descriptors are generated at the start with no additional expressions created during SIS.
Because of this, one MPI-task is used per GPU, with the remaining cores evenly distributed per task as defined by the machine using OpenMP.

\subsection{Single-node performance}
\begin{figure*}
    \centering
    \includegraphics{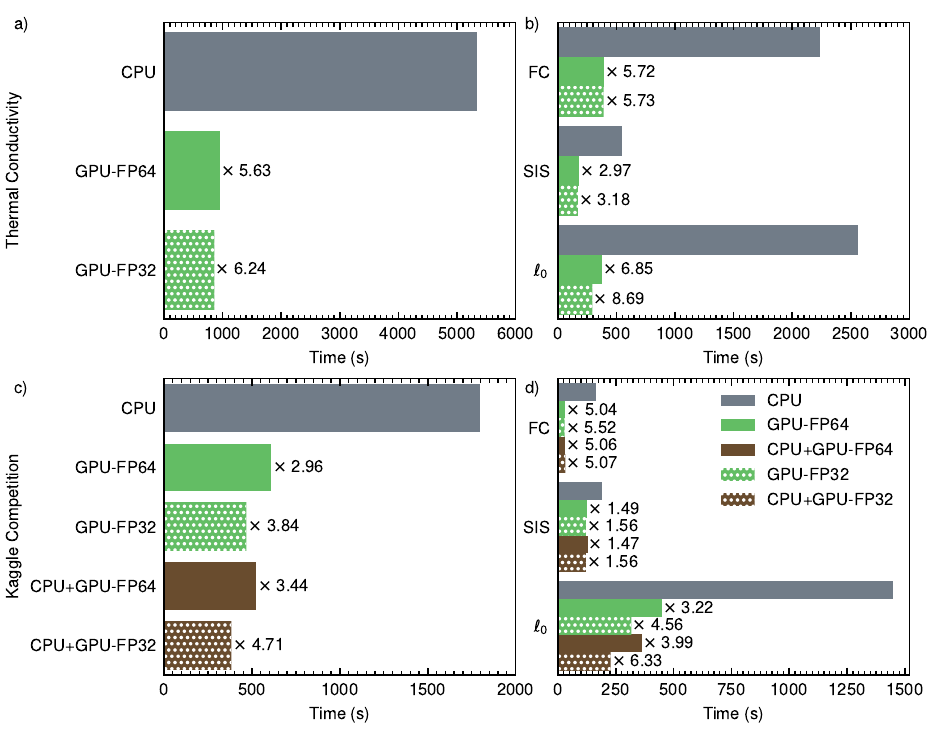}
    \caption{A comparison of the total run time (left column, panels a, c) with a breakdown into the major algorithmic parts (right column, panels b, d) for the thermal conductivity (top row, panels a, b) and Kaggle Competition Test Case (bottom row, panels c, d). The benchmark was executed on the Nvidia A100 platform (\textit{Raven}). The gray bar is drawn for CPU-only calculations (MKL-based baseline), the green bar represents GPU-enabled calculations in double precision, the brown bar is for a hybrid CPU-GPU setup for $\ell_0$ regularization, the ``x'' hatched bars are the corresponding results for single-precision (FP32). Bars involving GPU calculations are labelled by the obtained speedup relative to the CPU-only reference configuration, defined as the ratio $t_\mathrm{GPU} / t_\mathrm{CPU}$. The hybrid setup is not shown for a and b because those calculations cannot be efficiently run with more than one thread per MPI rank currently. }
    \label{fig:raven-singlenode}
\end{figure*}

Figure~\ref{fig:raven-singlenode} shows the single-node performance of the new Kokkos-enabled code, executed on a single node of the \textit{Raven} cluster using 4 Nvidia A100 GPUs and 72 Intel Xeon IceLake-SP CPU cores.
For the thermal conductivity test case (top row of Figure~\ref{fig:raven-singlenode}), we observe a significant improvement in the time-to-solution by a factor of 5.6 for the GPU-FP64 configuration, and by a factor of 6.2 for GPU-FP32, respectively, as compared to the CPU-only reference run (Figure~\ref{fig:raven-singlenode}a).
Note that for this test case GPU-FP64 and GPU-FP32 yield practically the same numerical results, making GPU-FP32 the fastest choice.
We also note that the obtained GPU speedups significantly overcompensate the economic cost of the GPUs (hardware investment for the GPUs plus additional electricity costs are estimated to make the node more expensive by a factor of three to four compared to a CPU-only configuration), making our new code economically viable, also in a commercial setting \cite{amazon2024}.

Looking at the breakdown of the individual code parts (FC, SIS, $\ell_0$) shown in Figure~\ref{fig:raven-singlenode}b this is largely a result of the acceleration obtained in the $\ell_0$-regularization.
Interestingly, switching from double precision (GPU-FP64) to single precision (GPU-FP32) does not further improve the performance of feature-creation or SIS steps considerably.
This is attributed to the building and maintenance of the expression tree and unit checks which now use the most of the compute time and cannot take advantage of a lower arithmetic precision.
However, contrary to expectations, the performance increase from FP64 to FP32 for $\ell_0$-regularization is also quite moderate. Here, the creation and ranking of the individual models, which does not involve floating point operations, becomes dominant due to the relatively small problem size (feature vectors only have $\approx 400$ entries).
The small problem size also makes the GPU kernels more integer-operation heavy (index calculation, loop counters, etc.).

For the larger Kaggle competition data-set (bottom row of Figure~\ref{fig:raven-singlenode}) we obtain an overall speed-up by a factor of 4.71 for the GPU-FP32 configuration (Figure~\ref{fig:raven-singlenode}c).
Also for this test case, FP32 yields the same numerical results as FP64.
The speed-up cannot be compared directly between the two test cases as the different problem sizes result in a different ratio between CPU and GPU workload as well as different overheads for caching and data structure handling.
The trend for the FC and SIS parts is the same as described above for the smaller data-set, while the $\ell_0$-regularization shows an expected, significant speed-up from FP64 to FP32 (Figure~\ref{fig:raven-singlenode}d), due to floating-point operations becoming more dominant for the larger problems size.
For this test case we are able to use the hybrid parallelization for the $\ell_0$-regularization, denoted by CPU+GPU-*. With the combined compute power of the whole system the performance is even better and we achieve a speed-up of up to 6.3.

\subsubsection{Multi-node scalability}

\begin{figure*}
    \centering
    \includegraphics{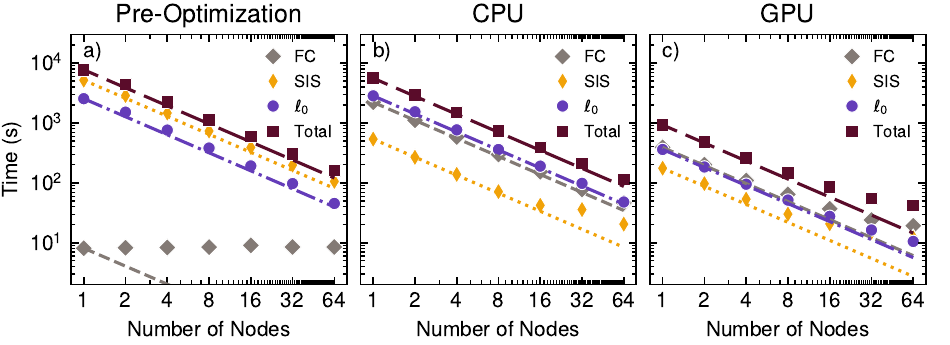}
    \caption{Runtime as a function of the number of compute nodes on the A100 platform (\textit{Raven}) with the thermal conductivity test case executed on a the a) pre-optimized version, b) CPU-only, and c) GPU-accelerated code. The total runtime (maroon squares) is shown together with a breakdown into the code parts FC (gray diamonds), SIS (gold diamonds), and $\ell_0$-regularization (purple circles). Ideal strong scaling is indicated by dashed lines.}
    \label{fig:mpi}
\end{figure*}

Figure~\ref{fig:mpi} shows the strong-scaling behaviour of the new code and its most relevant constituents on up to 64 nodes (4608 Xeon IceLake-SP CPU cores, 256 Nvidia A100 GPUs) of the HPC system \textit{Raven}. For this study we replicate the thermal conductivity test set on a series of different numbers of nodes. Despite the significant speedup of the single-node performance achieved by GPU-acceleration, the parallel scaling of the total runtime remains good for all parts of the code. During the porting of the code, the on-the-fly feature creation portions were recategorized as part of feature creation instead of SIS. This explains the sudden improvement seen in SIS and the increase in the FC line. Overall porting the code onto GPUs improved the overall performance of the algorithms even on CPUs as a result of general optimizations. Importantly the strong-scaling of original code is preserved up to the point where the serial-only portion of the code dominates. For feature creation this is illustrated by the pre-optimized feature creation step data that is constant at $\approx$10 s. During this time all of the generated features for a particular rung need to be gathered, checked to ensure none of the expressions can be simplified into already existing ones, and redistributed across all MPI ranks. For SIS similar delays as described for feature creation are encountered, while for $\ell_0$ regularized this is a few seconds to gather, write, and distribute the final selected models across all MPI-ranks.

\subsection{Performance Portability}
\begin{figure*}
    \centering
    \includegraphics{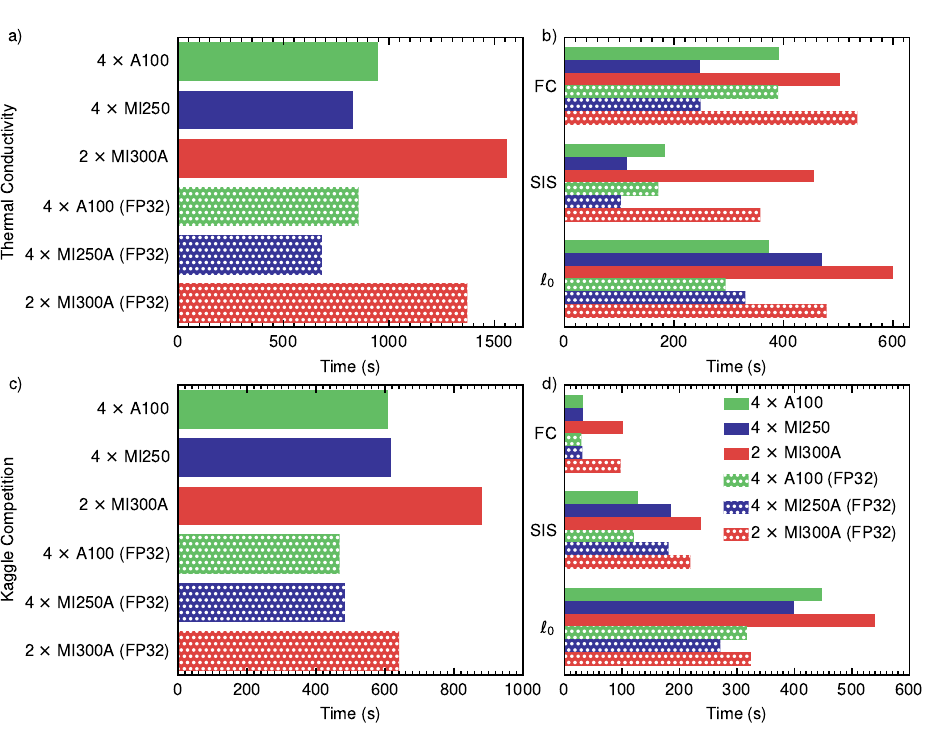}
    \caption{A comparison of the total run time (left column, panels a, c) with a breakdown into the major algorithmic parts (right column, panels b, d) for the thermal conductivity (top row, panels a, b) and $\ell_0$ (bottom row, panels c, d) benchmarks, executed on a single node with different hardware setups: The green, blue, and red bars show calculation times on  Nvidia A100 GPUs, and AMD MI250 and MI300A GPUs, respectively, and the ``x'' hatched bars show single-precision (FP32) results for the same platforms. For A100 and MI250 results there were 4 GPUs per node and MI300A has 2 GPUs per node. }
    \label{fig:AMD_porting}
\end{figure*}
After developing, testing, and optimizing the new code on the A100 platform, we performed additional benchmarks on GPU-accelerated platforms of a different vendor (AMD) and different hardware generations, including the recently introduced AMD MI300A. Thanks to our Kokkos port of SISSO++ this simply (and solely) required switching the Kokkos backend from CUDA (AMPERE80) to ROCm/HIP (gfx90a for the MI250, gfx942 for the MI300A). Notably, Kokkos was enabled for MI300A very early on (we started our experiments on AAC in May 2024), even before the hardware became generally available for developers and users.

Although, given the raw peak performances of the hardware (Table~\ref{tab:hardware})  the MI250 GPU (2 GCDs) should outperform one A100 GPU significantly, we observe a somewhat mixed picture for SISSO++ where our corresponding speedups range from 0.5x up to 1.5x depending on the kernel. By and large, this in accordance to what other scientists have observed for their HPC codes. \citet{Schlepphorst2023} report around 0.5x, \citet{Halbiniak2023} and \citet{Halver2023} see roughly equal performance, whereas \citet{Jimenez2022} and \citet{Thavappiragasam2024} measure up to twice the performance for MI250 (2 GCDs) compared to one A100 for their respective codes.

Similarly, with the new MI300A platform just entering the field, data for comparing and qualifying real-world application performance on MI300A is still very scarce at the time of writing \cite{tandon2024,ruzicka2024}. According to our own --- still very early --- experiences at MPCDF, complex HPC codes should expect a factor of between 1.5x and 2.5x per device, when comparing application performance on AMD's MI300A with Nvidia's A100, which is consistent with our SISSO++ benchmarks on MI300A reported here.

In-depth performance profiling and modelling is beyond the scope of this work, but clearly needs to be conducted in order to guide further optimization efforts of SISSO++ on the various (new) hardware platforms. A basic analysis showed that remaining bottlenecks are mostly located in parts of the code that are not yet executed on the GPU, indicating opportunities for further speeding up the code. In this respect, the new platforms AMD MI300A and Nvidia Grace-Hopper with their (physically) unified memories will be interesting targets of investigation, regarding their ability to tolerate even a fine-grained mix of CPU and GPU operations.

\section{Conclusion}
In this paper we reported about our efforts to make the SISSO++ code, which implements
an artificial intelligence method based on symbolic regression and compressed sensing for materials science research,
ready for current and future HPC platforms.
To tackle the problem of supporting multiple different hardware architectures with preferably a single code base, we adopted the Kokkos library.
After carefully assessing all data structures and algorithms employed by SISSO++, we converted the most time consuming parts of the code to Kokkos.
Overall, the required code changes to use Kokkos as the parallel backend for SISSO++ went smoothly and no major obstacles were encountered. The unified memory capabilities of AMD and Nvidia GPUs made it possible to transfer the code incrementally to the GPU. We showed the portability of our new Kokkos-based code between Intel and AMD CPUs and AMD and Nvidia GPUs,
without the need for hardware-dependent or vendor-specific code changes, thus confirming the main promise of the Kokkos framework.

For the evaluation of the performance, we chose two example setups.
One adapting a setup previously used to explore thermal insulators and one larger setup for exploring future prospects.
The adaptations made for the thermal insulator dataset was done to ensure roughly equal performance for all three components, and including computational data in a multi-task example.
Running on one compute node equipped with two Intel Xeon IceLake 8360Y CPUs (2x36 cores) and four Nvidia A100 GPUs we were able to achieve a GPU-speedup of 6.2x for the thermal insulator setup and 4.7x for the larger setup. This result exceeds the economic break even point for buying and running a GPU based compute node.
With single-precision calculations providing strong boost for the numerical parts of the code.
We also showed performance portability to AMD MI250 GPUs and the new MI300A APU. Our results are well in accordance with what other researchers have reported.

The MPI-based parallelization of the SISSO++ code stays untouched, and we showcased the strong-scaling capability of the code on up to 32 compute nodes (128 A100 GPUs), which enabled reducing the total runtime from 1.5 hours (the original, highly optimized CPU code executed on 72 CPU cores) to less than 4 minutes (the new Kokkos-enabled code using 128 A100 GPUs) for a scientifically relevant test case.

This advancement represents a significant step toward enabling the SISSO method for future more complex and larger applications and using the most modern HPC platforms.
With additional feature reduction strategies it might even be possible to extend the complexity of the candidate descriptor to the fourth rung, despite the combinatorial bottleneck.
Importantly, the acceleration of the software enables the generation of an ensemble of accurate symbolic regression models needed in active learning techniques.
Here active learning refers to using an AI model with uncertainty estimates to select which new data points should have their target properties evaluated next.
This cycle is then repeated sequentially, and by reducing the overall training time by close to an order of magnitude for real-world applications, SISSO is now able to train a large ensemble of models to estimate the uncertainty for each prediction.
SISSO++ is freely available (under the Apache 2 Licence) at \url{https://gitlab.com/sissopp_developers/sissopp}.

\section*{CRediT author statement}
Sebastian Eibl: Methodology, Software(l0), Writing

Yi Yao: Software(FC, SIS), Writing

Matthias Scheffler:Supervison

Markus Rampp: Supervision, Writing

Luca M. Ghiringhelli: Supervision

Thomas A. R. Purcell: Supervision, Software (FC, SIS), Visualization, Writing

\section*{Acknowledgements}
This work was funded by the NOMAD Center of Excellence (European Union's Horizon 2020 research and innovation program, grant agreement Nº 951786), the ERC Advanced Grant TEC1p (Eurpoean Research Council, Grant AGreement No. 740233) and BigMax (the Max Planck Society's Research Network on Big-Data-Driven Materials-Science). T.P. would like to thank the Alexander von Humboldt (AvH) Foundation for their support through the AvH Postdoctoral Fellowship Program.
We want to thank AMD for providing us access to their GPUs in the AMD Accelerator Cloud. In particular, early access to MI300A hardware is gratefully acknowledged.
We would also like to thank Markus Hrywniak (Nvidia) for sharing his expertise about Nvidia software tools and kernel optimizations.

\bibliographystyle{elsarticle-num-names}
\bibliography{literature}

\begin{thebibliography}{28}
\expandafter\ifx\csname natexlab\endcsname\relax\def\natexlab#1{#1}\fi
\providecommand{\url}[1]{\texttt{#1}}
\providecommand{\href}[2]{#2}
\providecommand{\path}[1]{#1}
\providecommand{\DOIprefix}{doi:}
\providecommand{\ArXivprefix}{arXiv:}
\providecommand{\URLprefix}{URL: }
\providecommand{\Pubmedprefix}{pmid:}
\providecommand{\doi}[1]{\href{http://dx.doi.org/#1}{\path{#1}}}
\providecommand{\Pubmed}[1]{\href{pmid:#1}{\path{#1}}}
\providecommand{\bibinfo}[2]{#2}
\ifx\xfnm\relax \def\xfnm[#1]{\unskip,\space#1}\fi
\bibitem[{TOP(2024)}]{TOP500}
\bibinfo{title}{{TOP500 list June 2024}},
  \bibinfo{howpublished}{\url{https://www.top500.org/lists/top500/2024/06/}},
  \bibinfo{year}{2024}. \bibinfo{note}{[Online; accessed 18-September-2024]}.
\bibitem[{Davis et~al.(2024)Davis, Sivaraman, Minn, Parasyris, Menon,
  Georgakoudis, and Bhatele}]{davis2024}
\bibinfo{author}{J.~H. Davis}, \bibinfo{author}{P.~Sivaraman},
  \bibinfo{author}{I.~Minn}, \bibinfo{author}{K.~Parasyris},
  \bibinfo{author}{H.~Menon}, \bibinfo{author}{G.~Georgakoudis},
  \bibinfo{author}{A.~Bhatele},
\newblock \bibinfo{title}{An evaluative comparison of performance portability
  across gpu programming models}  (\bibinfo{year}{2024}). \URLprefix
  \url{https://arxiv.org/abs/2402.08950v1}.
\bibitem[{Marowka(2022)}]{marowka2022}
\bibinfo{author}{A.~Marowka},
\newblock \bibinfo{title}{On the performance portability of openacc, openmp,
  kokkos and raja},
\newblock in: \bibinfo{booktitle}{International Conference on High Performance
  Computing in Asia-Pacific Region}, HPCAsia '22,
  \bibinfo{publisher}{Association for Computing Machinery},
  \bibinfo{address}{New York, NY, USA}, \bibinfo{year}{2022}, p.
  \bibinfo{pages}{103–114}. \URLprefix
  \url{https://doi.org/10.1145/3492805.3492806}.
  \DOIprefix\doi{10.1145/3492805.3492806}.
\bibitem[{Herten(2023)}]{herten2023}
\bibinfo{author}{A.~Herten},
\newblock \bibinfo{title}{Many cores, many models: Gpu programming model vs.
  vendor compatibility overview},
\newblock in: \bibinfo{booktitle}{Proceedings of the SC '23 Workshops of The
  International Conference on High Performance Computing, Network, Storage, and
  Analysis}, SC-W '23, \bibinfo{publisher}{Association for Computing
  Machinery}, \bibinfo{address}{New York, NY, USA}, \bibinfo{year}{2023}, p.
  \bibinfo{pages}{1019–1026}. \URLprefix
  \url{https://doi.org/10.1145/3624062.3624178}.
  \DOIprefix\doi{10.1145/3624062.3624178}.
\bibitem[{Latt et~al.(2021)Latt, Coreixas, and Beny}]{Latt2021}
\bibinfo{author}{J.~Latt}, \bibinfo{author}{C.~Coreixas},
  \bibinfo{author}{J.~Beny},
\newblock \bibinfo{title}{Cross-platform programming model for many-core
  lattice boltzmann simulations},
\newblock \bibinfo{journal}{PLOS ONE} \bibinfo{volume}{16}
  (\bibinfo{year}{2021}) \bibinfo{pages}{e0250306}.
  \DOIprefix\doi{10.1371/journal.pone.0250306}.
\bibitem[{Beckingsale et~al.(2019)Beckingsale, Scogland, Burmark, Hornung,
  Jones, Killian, Kunen, Pearce, Robinson, and Ryujin}]{Beckingsale2019}
\bibinfo{author}{D.~A. Beckingsale}, \bibinfo{author}{T.~R. Scogland},
  \bibinfo{author}{J.~Burmark}, \bibinfo{author}{R.~Hornung},
  \bibinfo{author}{H.~Jones}, \bibinfo{author}{W.~Killian},
  \bibinfo{author}{A.~J. Kunen}, \bibinfo{author}{O.~Pearce},
  \bibinfo{author}{P.~Robinson}, \bibinfo{author}{B.~S. Ryujin},
\newblock \bibinfo{title}{Raja: Portable performance for large-scale scientific
  applications},
\newblock in: \bibinfo{booktitle}{2019 IEEE/ACM International Workshop on
  Performance, Portability and Productivity in HPC (P3HPC)},
  \bibinfo{publisher}{IEEE}, \bibinfo{year}{2019}.
  \DOIprefix\doi{10.1109/p3hpc49587.2019.00012}.
\bibitem[{Trott et~al.(2022)Trott, Lebrun-Grandié, Arndt, Ciesko, Dang,
  Ellingwood, Gayatri, Harvey, Hollman, Ibanez, Liber, Madsen, Miles,
  Poliakoff, Powell, Rajamanickam, Simberg, Sunderland, Turcksin, and
  Wilke}]{Trott2022}
\bibinfo{author}{C.~R. Trott}, \bibinfo{author}{D.~Lebrun-Grandié},
  \bibinfo{author}{D.~Arndt}, \bibinfo{author}{J.~Ciesko},
  \bibinfo{author}{V.~Dang}, \bibinfo{author}{N.~Ellingwood},
  \bibinfo{author}{R.~Gayatri}, \bibinfo{author}{E.~Harvey},
  \bibinfo{author}{D.~S. Hollman}, \bibinfo{author}{D.~Ibanez},
  \bibinfo{author}{N.~Liber}, \bibinfo{author}{J.~Madsen},
  \bibinfo{author}{J.~Miles}, \bibinfo{author}{D.~Poliakoff},
  \bibinfo{author}{A.~Powell}, \bibinfo{author}{S.~Rajamanickam},
  \bibinfo{author}{M.~Simberg}, \bibinfo{author}{D.~Sunderland},
  \bibinfo{author}{B.~Turcksin}, \bibinfo{author}{J.~Wilke},
\newblock \bibinfo{title}{Kokkos 3: Programming model extensions for the
  exascale era},
\newblock \bibinfo{journal}{IEEE Transactions on Parallel and Distributed
  Systems} \bibinfo{volume}{33} (\bibinfo{year}{2022})
  \bibinfo{pages}{805--817}. \DOIprefix\doi{10.1109/TPDS.2021.3097283}.
\bibitem[{Ruzicka et~al.(2024)Ruzicka, Asch, Meneses, Rampp, and
  Laure}]{ruzicka2024}
\bibinfo{author}{J.~Ruzicka}, \bibinfo{author}{C.~Asch},
  \bibinfo{author}{E.~Meneses}, \bibinfo{author}{M.~Rampp},
  \bibinfo{author}{E.~Laure},
\newblock \bibinfo{title}{A study of performance portability in plasma physics
  simulations},
\newblock in: \bibinfo{booktitle}{CARLA 2024}, \bibinfo{year}{2024}.
\bibitem[{Purcell et~al.(2022)Purcell, Scheffler, Carbogno, and
  Ghiringhelli}]{Purcell2022}
\bibinfo{author}{T.~A.~R. Purcell}, \bibinfo{author}{M.~Scheffler},
  \bibinfo{author}{C.~Carbogno}, \bibinfo{author}{L.~M. Ghiringhelli},
\newblock \bibinfo{title}{{SISSO++: A C++ Implementation of the
  Sure-Independence Screening and Sparsifying Operator Approach}},
\newblock \bibinfo{journal}{J. Open Source Softw.} \bibinfo{volume}{7}
  (\bibinfo{year}{2022}) \bibinfo{pages}{3960}. \URLprefix
  \url{https://doi.org/10.21105/joss.03960.
  https://joss.theoj.org/papers/10.21105/joss.03960}.
  \DOIprefix\doi{10.21105/joss.03960}.
\bibitem[{Purcell et~al.(2023)Purcell, Scheffler, and
  Ghiringhelli}]{Purcell2023}
\bibinfo{author}{T.~A. Purcell}, \bibinfo{author}{M.~Scheffler},
  \bibinfo{author}{L.~M. Ghiringhelli},
\newblock \bibinfo{title}{Recent advances in the sisso method and their
  implementation in the sisso++ code},
\newblock \bibinfo{journal}{The Journal of Chemical Physics}
  \bibinfo{volume}{159} (\bibinfo{year}{2023}).
\bibitem[{Ouyang et~al.(2018)Ouyang, Curtarolo, Ahmetcik, Scheffler, and
  Ghiringhelli}]{Ouyang2017}
\bibinfo{author}{R.~Ouyang}, \bibinfo{author}{S.~Curtarolo},
  \bibinfo{author}{E.~Ahmetcik}, \bibinfo{author}{M.~Scheffler},
  \bibinfo{author}{L.~M. Ghiringhelli},
\newblock \bibinfo{title}{{SISSO: A compressed-sensing method for identifying
  the best low-dimensional descriptor in an immensity of offered candidates}},
\newblock \bibinfo{journal}{Phys. Rev. Mater.} \bibinfo{volume}{2}
  (\bibinfo{year}{2018}) \bibinfo{pages}{83802}.
\bibitem[{Ouyang et~al.(2019)Ouyang, Ahmetcik, Carbogno, Scheffler, and
  Ghiringhelli}]{Ouyang2019}
\bibinfo{author}{R.~Ouyang}, \bibinfo{author}{E.~Ahmetcik},
  \bibinfo{author}{C.~Carbogno}, \bibinfo{author}{M.~Scheffler},
  \bibinfo{author}{L.~M. Ghiringhelli},
\newblock \bibinfo{title}{{Simultaneous learning of several materials
  properties from incomplete databases with multi-task SISSO}},
\newblock \bibinfo{journal}{J. Phys. Mater.} \bibinfo{volume}{2}
  (\bibinfo{year}{2019}) \bibinfo{pages}{24002}.
\bibitem[{Fan and Lv(2008)}]{Fan2008}
\bibinfo{author}{J.~Fan}, \bibinfo{author}{J.~Lv},
\newblock \bibinfo{title}{{Sure independence screening for ultrahigh
  dimensional feature space}},
\newblock \bibinfo{journal}{J. R. Stat. Soc. Ser. B Stat. Methodol.}
  \bibinfo{volume}{70} (\bibinfo{year}{2008}) \bibinfo{pages}{849--911}.
  \URLprefix \url{http://doi.wiley.com/10.1111/j.1467-9868.2008.00674.x}.
  \DOIprefix\doi{10.1111/j.1467-9868.2008.00674.x}.
  \href{http://arxiv.org/abs/0612857}{{\tt arXiv:0612857}}.
\bibitem[{Artigues et~al.(2020)Artigues, Kormann, Rampp, and
  Reuter}]{artigues2020}
\bibinfo{author}{V.~Artigues}, \bibinfo{author}{K.~Kormann},
  \bibinfo{author}{M.~Rampp}, \bibinfo{author}{K.~Reuter},
\newblock \bibinfo{title}{Evaluation of performance portability frameworks for
  the implementation of a particle-in-cell code},
\newblock \bibinfo{journal}{Concurrency and Computation: Practice and
  Experience} \bibinfo{volume}{32} (\bibinfo{year}{2020})
  \bibinfo{pages}{e5640}. \DOIprefix\doi{https://doi.org/10.1002/cpe.5640},
  \bibinfo{note}{e5640 cpe.5640}.
\bibitem[{Schild et~al.(2024)Schild, Räth, Eibl, Hallatschek, and
  Kormann}]{schikd2024}
\bibinfo{author}{N.~Schild}, \bibinfo{author}{M.~Räth},
  \bibinfo{author}{S.~Eibl}, \bibinfo{author}{K.~Hallatschek},
  \bibinfo{author}{K.~Kormann},
\newblock \bibinfo{title}{A performance portable implementation of the
  semi-lagrangian algorithm in six dimensions},
\newblock \bibinfo{journal}{Computer Physics Communications}
  \bibinfo{volume}{295} (\bibinfo{year}{2024}) \bibinfo{pages}{108973}.
  \URLprefix
  \url{https://www.sciencedirect.com/science/article/pii/S0010465523003181}.
  \DOIprefix\doi{https://doi.org/10.1016/j.cpc.2023.108973}.
\bibitem[{Evans et~al.(2022)Evans, Siegel, Draeger, Deslippe, Francois,
  Germann, Hart, and Martin}]{evans2022}
\bibinfo{author}{T.~M. Evans}, \bibinfo{author}{A.~Siegel},
  \bibinfo{author}{E.~W. Draeger}, \bibinfo{author}{J.~Deslippe},
  \bibinfo{author}{M.~M. Francois}, \bibinfo{author}{T.~C. Germann},
  \bibinfo{author}{W.~E. Hart}, \bibinfo{author}{D.~F. Martin},
\newblock \bibinfo{title}{A survey of software implementations used by
  application codes in the exascale computing project},
\newblock \bibinfo{journal}{The International Journal of High Performance
  Computing Applications} \bibinfo{volume}{36} (\bibinfo{year}{2022})
  \bibinfo{pages}{5--12}. \DOIprefix\doi{10.1177/10943420211028940}.
\bibitem[{Lesur et~al.(2023)Lesur, Baghdadi, Wafflard-Fernandez, Mauxion,
  Robert, and den Bossche}]{lesur2023}
\bibinfo{author}{G.~R.~J. Lesur}, \bibinfo{author}{S.~Baghdadi},
  \bibinfo{author}{G.~Wafflard-Fernandez}, \bibinfo{author}{J.~Mauxion},
  \bibinfo{author}{C.~M.~T. Robert}, \bibinfo{author}{M.~V. den Bossche},
\newblock \bibinfo{title}{Idefix: a versatile performance-portable godunov code
  for astrophysical flows}  (\bibinfo{year}{2023}). \URLprefix
  \url{https://arxiv.org/abs/2304.13746v1}.
\bibitem[{Roussel-Hard et~al.(2024)Roussel-Hard, Audit, Dessart, Padioleau, and
  Wang}]{roussel2024}
\bibinfo{author}{L.~Roussel-Hard}, \bibinfo{author}{{\'E}.~Audit},
  \bibinfo{author}{L.~Dessart}, \bibinfo{author}{T.~Padioleau},
  \bibinfo{author}{Y.~Wang},
\newblock \bibinfo{title}{Tackling exascale systems for astrophysics},
\newblock \bibinfo{journal}{Journal of Physics: Conference Series}
  \bibinfo{volume}{2742} (\bibinfo{year}{2024}). \URLprefix
  \url{https://api.semanticscholar.org/CorpusID:269320787}.
\bibitem[{Purcell et~al.(2023)Purcell, Scheffler, Ghiringhelli, and
  Carbogno}]{Purcell2023a}
\bibinfo{author}{T.~A. Purcell}, \bibinfo{author}{M.~Scheffler},
  \bibinfo{author}{L.~M. Ghiringhelli}, \bibinfo{author}{C.~Carbogno},
\newblock \bibinfo{title}{Accelerating materials-space exploration for thermal
  insulators by mapping materials properties via artificial intelligence},
\newblock \bibinfo{journal}{npj computational materials} \bibinfo{volume}{9}
  (\bibinfo{year}{2023}) \bibinfo{pages}{112}.
\bibitem[{Knoop et~al.(2023)Knoop, Purcell, Scheffler, and
  Carbogno}]{Knoop2023}
\bibinfo{author}{F.~Knoop}, \bibinfo{author}{T.~A. Purcell},
  \bibinfo{author}{M.~Scheffler}, \bibinfo{author}{C.~Carbogno},
\newblock \bibinfo{title}{Anharmonicity in thermal insulators: An analysis from
  first principles},
\newblock \bibinfo{journal}{Physical Review Letters} \bibinfo{volume}{130}
  (\bibinfo{year}{2023}) \bibinfo{pages}{236301}.
\bibitem[{Sutton et~al.(2019)Sutton, Ghiringhelli, Yamamoto, Lysogorskiy,
  Blumenthal, Hammerschmidt, Golebiowski, Liu, Ziletti, and
  Scheffler}]{Sutton2019}
\bibinfo{author}{C.~Sutton}, \bibinfo{author}{L.~M. Ghiringhelli},
  \bibinfo{author}{T.~Yamamoto}, \bibinfo{author}{Y.~Lysogorskiy},
  \bibinfo{author}{L.~Blumenthal}, \bibinfo{author}{T.~Hammerschmidt},
  \bibinfo{author}{J.~R. Golebiowski}, \bibinfo{author}{X.~Liu},
  \bibinfo{author}{A.~Ziletti}, \bibinfo{author}{M.~Scheffler},
\newblock \bibinfo{title}{Crowd-sourcing materials-science challenges with the
  nomad 2018 kaggle competition},
\newblock \bibinfo{journal}{npj Computational Materials} \bibinfo{volume}{5}
  (\bibinfo{year}{2019}) \bibinfo{pages}{111}.
\bibitem[{Amazon(2024)}]{amazon2024}
\bibinfo{author}{Amazon}, \bibinfo{title}{Compute savings plans for amazon ec2,
  22 june 2024}, \bibinfo{year}{5 Oct 2024}.
  \bibinfo{note}{Https://aws.amazon.com/savingsplans/compute-pricing/}.
\bibitem[{Schlepphorst and Krieg(2023)}]{Schlepphorst2023}
\bibinfo{author}{S.~Schlepphorst}, \bibinfo{author}{S.~Krieg},
\newblock \bibinfo{title}{Benchmarking a portable lattice quantum
  chromodynamics kernel written in kokkos and mpi},
\newblock in: \bibinfo{booktitle}{Proceedings of the SC ’23 Workshops of The
  International Conference on High Performance Computing, Network, Storage, and
  Analysis}, SC-W 2023, \bibinfo{publisher}{ACM}, \bibinfo{year}{2023}, pp.
  \bibinfo{pages}{1027--1037}. \DOIprefix\doi{10.1145/3624062.3624179}.
\bibitem[{Halbiniak et~al.(2023)Halbiniak, Meyer, and Rojek}]{Halbiniak2023}
\bibinfo{author}{K.~Halbiniak}, \bibinfo{author}{N.~Meyer},
  \bibinfo{author}{K.~Rojek},
\newblock \bibinfo{title}{Single‐ and multi‐gpu computing on nvidia‐ and
  amd‐based server platforms for solidification modeling application},
\newblock \bibinfo{journal}{Concurrency and Computation: Practice and
  Experience} \bibinfo{volume}{36} (\bibinfo{year}{2023}).
  \DOIprefix\doi{10.1002/cpe.8000}.
\bibitem[{Halver et~al.(2023)Halver, Junghans, and Sutmann}]{Halver2023}
\bibinfo{author}{R.~Halver}, \bibinfo{author}{C.~Junghans},
  \bibinfo{author}{G.~Sutmann},
\newblock \bibinfo{title}{Using heterogeneous gpu nodes with a cabana-based
  implementation of mpcd},
\newblock \bibinfo{journal}{Parallel Computing} \bibinfo{volume}{117}
  (\bibinfo{year}{2023}) \bibinfo{pages}{103033}.
  \DOIprefix\doi{10.1016/j.parco.2023.103033}.
\bibitem[{Jiménez et~al.(2022)Jiménez, Herrera-Mora, Rampp, Laure, and
  Meneses}]{Jimenez2022}
\bibinfo{author}{D.~Jiménez}, \bibinfo{author}{J.~Herrera-Mora},
  \bibinfo{author}{M.~Rampp}, \bibinfo{author}{E.~Laure},
  \bibinfo{author}{E.~Meneses}, \bibinfo{title}{Implementing a GPU-Portable
  Field Line Tracing Application with OpenMP Offload},
  \bibinfo{publisher}{Springer International Publishing}, \bibinfo{year}{2022},
  pp. \bibinfo{pages}{31--46}. \DOIprefix\doi{10.1007/978-3-031-23821-5_3}.
\bibitem[{Thavappiragasam et~al.(2024)Thavappiragasam, Harris, Endeve, and
  Videau}]{Thavappiragasam2024}
\bibinfo{author}{M.~Thavappiragasam}, \bibinfo{author}{J.~A. Harris},
  \bibinfo{author}{E.~Endeve}, \bibinfo{author}{B.~Videau},
  \bibinfo{title}{Performance Porting the ExaStar Multi-Physics App Thornado
  On Heterogeneous Systems - A Fortran-OpenMP Code-Base Evaluation},
  \bibinfo{publisher}{Springer Nature Switzerland}, \bibinfo{year}{2024}, pp.
  \bibinfo{pages}{16--30}. \DOIprefix\doi{10.1007/978-3-031-72567-8_2}.
\bibitem[{Tandon et~al.(2024)Tandon, Grinberg, Bercea, Bertolli, Olesen, Bna,
  and Malaya}]{tandon2024}
\bibinfo{author}{S.~Tandon}, \bibinfo{author}{L.~Grinberg},
  \bibinfo{author}{G.-T. Bercea}, \bibinfo{author}{C.~Bertolli},
  \bibinfo{author}{M.~Olesen}, \bibinfo{author}{S.~Bna},
  \bibinfo{author}{N.~Malaya},
\newblock \bibinfo{title}{Porting hpc applications to amd instinct™ mi300a
  using unified memory and openmp®},
\newblock in: \bibinfo{booktitle}{ISC High Performance 2024 Research Paper
  Proceedings (39th International Conference)}, \bibinfo{year}{2024}, pp.
  \bibinfo{pages}{1--9}. \DOIprefix\doi{10.23919/ISC.2024.10528925}.

\end{thebibliography}


\end{document}